\documentclass[
  ,draft            
  ]
  {aipproc}

\layoutstyle{6x9}

\begin{document}

\title{Thermonuclear supernova models, and observations of Type Ia supernovae}

\author{E. Bravo}{
  address={Dept. F\'\i sica i Enginyeria Nuclear, UPC, Av. Diagonal 647, 08028
  Barcelona},
  altaddress={Institut d'Estudis Espacials de Catalunya, Barcelona}
}

\author{C. Badenes}{
  address={Dept. Physics and Astronomy, Rutgers Univ., 136 Frelinghuysen Rd.,
  Piscataway NJ 08854-8019}
}

\author{D. Garc\'\i a-Senz}{
  address={Dept. F\'\i sica i Enginyeria Nuclear, UPC, Av. Diagonal 647, 08028
  Barcelona},
  altaddress={Institut d'Estudis Espacials de Catalunya, Barcelona}
}

\begin{abstract}
In this paper, we review the present state of theoretical models of 
thermonuclear supernovae, and compare their predicitions with the constraints
derived from observations of Type Ia supernovae. The
diversity of explosion mechanisms usually found in one-dimensional simulations 
is a direct consequence of the impossibility to 
resolve the flame structure under the assumption of spherical symmetry. 
Spherically symmetric models have been successful
in explaining many of the observational features of Type Ia
supernovae, but they rely on two kinds of empirical models: one that describes
the behaviour of the flame on the scales unresolved by the code, and another that
takes account of the evolution of the flame shape. In contrast,
three-dimensional simulations are able to compute the
flame shape in a self-consistent way, but they still need a model for the
propagation of the flame in the scales unresolved by the code. Furthermore, in 
three dimensions the
number of degrees of freedom of the initial configuration of the white dwarf at
runaway is much larger than in one dimension. Recent simulations have shown
that the sensitivity of the explosion output to the initial conditions can be
extremely large. New paradigms of
thermonuclear supernovae have emerged from this situacion, as the Pulsating
Reverse Detonation. The resolution of all these issues must rely on the predictions of observational
properties of the models, and their comparison with current Type Ia supernova data, including
X-ray spectra of Type Ia supernova remnants.
\end{abstract}

\maketitle


\section{Introduction}

The huge increase in number, quality and diversity of observational data 
related to Type Ia supernovae (SNIa) in recent years, 
combined with the advance in computer technology, have persuaded modellers
to leave the phenomenological calculations that rely on   
spherical symmetry, and attempt more physically meaningful 
three-dimensional (3D) simulations. Although the more plausible models of the 
explosion
always involve the thermonuclear disruption of a white dwarf
\citep{HoyleFowler1960}, the current zoo 
of explosion mechanisms is still
too large to be useful in cosmological applications of Type Ia supernovae
 or to make it possible to understand the details of the chemical evolution
of the Galaxy. Nowadays, the favoured SNIa
model is the explosion of a white dwarf that approaches the 
Chandrasekhar-mass limit owing to accretion from a companion star at the 
appropiate rate to avoid the nova instability
\citep{Nugent1997,HillebrandtNiemeyer2000}. Going beyond this general picture
into the details of the supernova explosion is not easy, especially with respect
to the
multidimensional models that are just beginning to appear in the literature. In
particular, the prediction of the optical light curve or spectra of a 3D model 
is still out of reach, and therefore it is 
necessary to rely on other gross features of the observations 
in order to estimate the viability of a given model.

The most relevant property of SNIa is the homogeneity of their light curve 
and spectral
evolution. The light curve is powered by the radioactive decay of $^{56}$Ni and 
$^{56}$Co \citep{ColgateMcKee1969}, but the range of nickel masses allowed by
the observations varies by about a factor five from low-luminosity SNIa up to
normal events. Although the shape of the light curves can be described 
by a one-parameter relationship between brightness and width of
the curve \citep{Riess1996,Hamuy1996,Perlmutter1997,Phillips1999}, due to 
the dependence of opacity on
temperature, there still remains a residual scatter of 
$\sim0.2$~mag around the template curves. The main spectral features of normal
(bright) SNIa at early photospheric phase include the absence of conspicuous
lines of H and the presence of strong SiII absorption lines together with
absorption lines of other intermediate mass elements (CaII, SII, OI) spanning a
range of velocities from 8,000 up to 30,000~km\,s$^{-1}$. The nebular phase is
dominated by Fe lines. Usually, the spectral evolution is attributed to the
recession (in terms of lagrangian mass) of the photosphere through a layered 
chemical structure. Recent spectroscopic observations
of a dozen  {\sl Branch-normal} Type Ia supernovae in the near infrared 
\citep{Marion2003} suggest 
that the unburnt matter ejected has to be less than $10\%$~of the mass of the 
progenitor. According to these results, the presence of a substantial 
amount of unburnt low-velocity 
carbon near the center of the star is rather improbable. 

A relevant question for multidimensional simulations, is whether there is 
any significant observational evidence of departure from spherical symmetry in
the SNIa sample. 
In this regard, there are several signs that the departure from spherical 
symmetry is not 
large: the low level of polarization of most SNIa, 
although there are exceptions \citep[see, for instance,][]{Kasen2003}, 
the homogeneity of the profile of 
the absortion line of SiII \citep{Thomas2002}, and the fact that galactic and
extra-galactic young Type Ia supernova 
remnants (SNR) do not show large departures from spherical symmetry. 

The spectral homogeneity of normal SNIa near maximum 
brightness is particularly relevant for the discussion below. 
By comparing the 
spectra of four normal SNIa (SN 1989B,
SN 1990N, SN 1994D, and SN 1998bu) \citet{Thomas2002} have shown that the 
absorption
features of SiII displayed quite homogeneous profiles from event to event.
Such homogeneity can be used to constrain the presence of chemically 
inhomogeneous clumps  
at the photosphere, through the effect they would have on the line profiles. 
Specifically, \citet{Thomas2002} limited the size of the clumps to be less than
$\sim30\%$ of the radius of the photosphere.  

To summarize the gross constraints imposed by SNIa observations, 
the {\sl overall} shape of the supernova has to be spherical (low
polarization), there are not large chemically inhomogeneous blobs at the
photosphere at maximum light (homogeneous SiII line profiles), and the chemical
composition of the ejecta has to retain a high degree of stratification. 
One has to keep in
mind that subluminous as well as superluminous events display a peculiar spectral
evolution. For the time being, however, as long as multidimensional simulations
are concerned, the main objective is to explain the gross properties of
normal SNIa.
In the first part of this paper, we will review the results of recent 3D 
simulations of thermonuclear
supernovae (TSN), and compare their predictions with observational data.
In the second part, we will discuss the prospects for the use of X-ray spectra 
of supernova	
remnants to discriminate between the different explosion mechanisms
or progenitor scenarios that are currently advocated to explain SNIa.

\section{3D models of thermonuclear supernovae}

From a theoretical point of view, there remain several fundamental
issues to be solved:
\begin{itemize}
\item What is the progenitor system and how does the white dwarf manage to
reach the Chandrasekhar mass?
\item What is the evolution of the white dwarf short before carbon runaway? and
how, when, and where does the ignition process begin?
\item How does the flame propagate through the white dwarf once ignited? 
\item Is a deflagration-to-detonation transition possible, and under
which conditions?
\item What is the role played by the rotation of the white dwarf?
\end{itemize}

Here, we will not discuss the presupernova evolution,
which is addressed at length in other contributions to this volume. However, it
is worth to remark that the output of the explosion in terms of its kinetic
energy, density profile, and chemical composition is determined in the first
few seconds after runaway, and is not much sensitive to
the details of the presupernova evolution. The only features of the
progenitor system that influence the explosion are the C/O ratio and metallicity
of the white dwarf, its central density (determined by the accretion rate), and
rotation (see the contributions by Piersanti et al. and Dom\'\i nguez et al. in
this volume). However, the supernova properties can be influenced on longer 
timescales by several characteristics of the progenitor system, owing to the 
interaction of the ejecta with the secondary star (this is probably the case for
the peculiar supernovae SN2000cx and SN2002cx, see e.g. 
\citet{Thomas2003,Li2003}), with a circumstellar medium (normal Type
Ia SN2003du and SN2002ic, see \citet{Gerardy2004,Hamuy2003}), or during the 
formation of the supernova remnant \citep{Badenes2004}. 

Spherically symmetric models and early two and three-dimensional
calculations of TSN assumed that the ignition started in a central volume.
This view was challenged by \citet{GarciaWoosley1995}, who showed that
burning blobs formed during the convective preignition phase would be able to
float and accelerate up to $\sim100$~ km\,s$^{-1}$, with the result that the
flame would be rapidly scattered in a region $100-250$~km away from the center 
of the
star. In this case, the flame would not begin just in a central volume but 
distributed
in an exponentially increasing \citep{Woosley2004,GarciaBravo2004} number of 
hot spots, whose velocity could reach $\sim1\%$ of the sound velocity. 

In order to simulate TSN it is always necessary to approximate the behavior of
the flame below the scales not resolved by the hydrocode with a suitable
model. This is
not an easy task, due to the quite different regimes of thermonuclear flames at
high densities, $\rho\sim10^9$~g\,cm$^{-3}$, (thin flame of width $< 1$~cm, 
propagating with a low velocity of
order $1-3\%$ of the sound velocity, and with a surface progressively corrugated
by hydrodynamic instabilities on timescales of a few tenths of a second) 
and at the 
end of the explosion, $\rho <$ a few $\times10^7$~g\,cm$^{-3}$, (thick flame 
of width
similar to the white dwarf radius, subject to mixing between ashes and fuel {\sl
before} completion of the nuclear reactions, which favours the production of 
intermediate mass elements). The range of involved lengthscales spans $\sim9$
orders of magnitude, which on one hand discards its direct resolution with any
hydrodynamical code but, on the other hand, allows to use a statistical 
description of the flame. Up to now, there is no convergence between the 
different approximations adopted by different 3D hydrocodes. 

\subsection{Deflagrations and delayed detonations}

The multidimensional calculations of deflagrations carried out so far 
\citep[see][for the most recent results]
{Reinecke2002,Gamezo2003,
GarciaBravo2004}
have shown interesting 
deviations from what is predicted in spherically symmetric models: 
\begin{enumerate}
\item The   
geometry of the burning front is no longer spherical owing to the important 
role played by buoyancy and hydrodynamic instabilities, 
\item the chemical stratification of the ejecta is lost,
\item the amount of $^{56}$Ni is sufficient to power the light curve, 
but it is localized 
in clumps distributed all along 
the radius of the white dwarf, and 
\item an uncomfortably large amount of carbon and 
oxygen remains unburnt at the center of the white dwarf. 
\end{enumerate}
Three-dimensional simulations have also demonstrated that the flame evolves in
quite a different way when calculated in 2D or 3D, due to the different degrees
of freedom of the flow. Thus, earlier results of 2D models of thermonuclear
supernovae have to be regarded with caution.

The results of the most up-to-date 3D simulations of deflagration supernovae 
(kinetic energy, $K$, and
masses of $^{56}$Ni, $M_{56}$, intermediate mass elements, $M_\mathrm{ime}$, and
unburned C+O, $M_\mathrm{CO}$) are shown in Table~\ref{tab:a}. The results
obtained under quite different initial conditions and using very different
numerical techniques (PPM vs SPH, with degrees of spatial resolution varying by
a factor ten, with different subgrid-scale models of the flame, etc) are
remarkably 
homogeneous. The kinetic energy and the amount of $^{56}$Ni are compatible with
SNIa, but the amount of intermediate mass elements is rather low and
 the mass of C+O ejected in the explosion is too large, probably by a
factor five or more. The convergence of the results obtained with different
codes reflects the fact that the main trends of deflagration supernovae are well
understood and incorporated into the calculations. 
There is little hope that further refinements in
the methods used to simulate them will substantially change the outcome of
current 3D deflagration models.

\begin{table}
\begin{tabular}{lcccc}
\hline
   \tablehead{1}{c}{b}{Model}
  & \tablehead{1}{c}{b}{$K$\\($10^{51}$~erg)}
  & \tablehead{1}{c}{b}{$M_{56}$\\($M_{\odot}$)}
  & \tablehead{1}{c}{b}{$M_\mathrm{ime}$\\($M_{\odot}$)}
  & \tablehead{1}{c}{b}{$M_\mathrm{CO}$\\($M_{\odot}$)}
  \\
\hline
{\bf Deflagrations} & & & & \\
\hline
Central ignition with strong turbulence\citep{Gamezo2003} & 
0.6 & $\sim0.5$ & $\sim0.1$ & $\sim0.7$ \\
Central ignition with realistic turbulence\citep{Gamezo2003} & 
0.37 & & & \\
Central ignition\citep{Reinecke2002} & 
0.48 & 0.30 & 0.10 & 0.75 \\
Ignition in 40 bubbles, large resolution\citep{Reinecke2002} & 0.45 & 0.33 & 0.23 & 0.64 \\
Ignition in 30 bubbles of the same size\citep{GarciaBravo2004}
& 0.43 & 0.43 & 0.07 & 0.67 \\
Ignition in 90 bubbles of different sizes\citep{GarciaBravo2004} & 0.45 & 0.44 & 0.08 & 0.66 \\
Ignition in 240 bubbles, very large resolution\citep{Niemeyer2003}
& 0.6 & 0.42 & 0.10 & 0.62 \\
\hline
{\bf Delayed detonations} & & & & \\
\hline
Macroscopic transition to detonation\citep{GarciaBravo2002} 
& 0.75 & 0.54 & 0.16 & 0.34 \\
Local transition in the central region\citep{GarciaBravo2002} & 0.48 & 0.43 & 0.10 & 0.48 \\
Local transition at intermediate radius\citep{GarciaBravo2002} & 0.51 & 0.42 & 0.14 & 0.45 \\
Local transition in the outer layers\citep{GarciaBravo2002} & 0.33 & 0.34 & 0.09 & 0.57 \\
Transition in a central volume\citep{Gamezo2004} & 
0.8 & 0.78\tablenote{In this and the following models $M_{56}$ represents the
approximate yield of Fe-group nuclei} & & \\
Off-center transition\citep{Gamezo2004} & 0.8 & 0.73 & & \\
Transition in a central volume at high density\citep{Gamezo2004} & 1.1 & 0.94 & & \\
\hline
\end{tabular}
\caption{Results of 3D simulations of thermonuclear supernovae}
\label{tab:a}
\end{table}

\begin{figure}
  \includegraphics[height=.33\textheight]{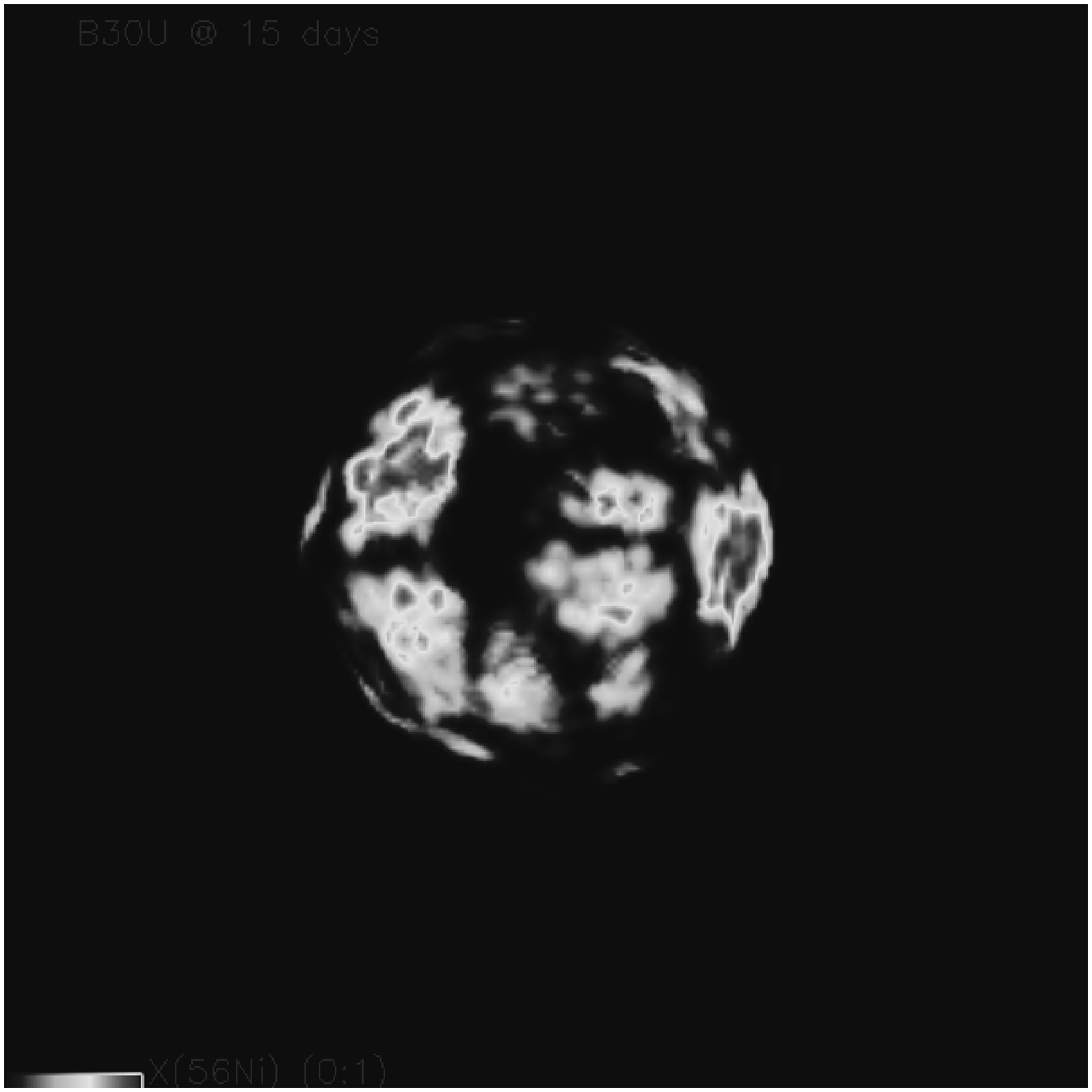}
  \includegraphics[height=.33\textheight]{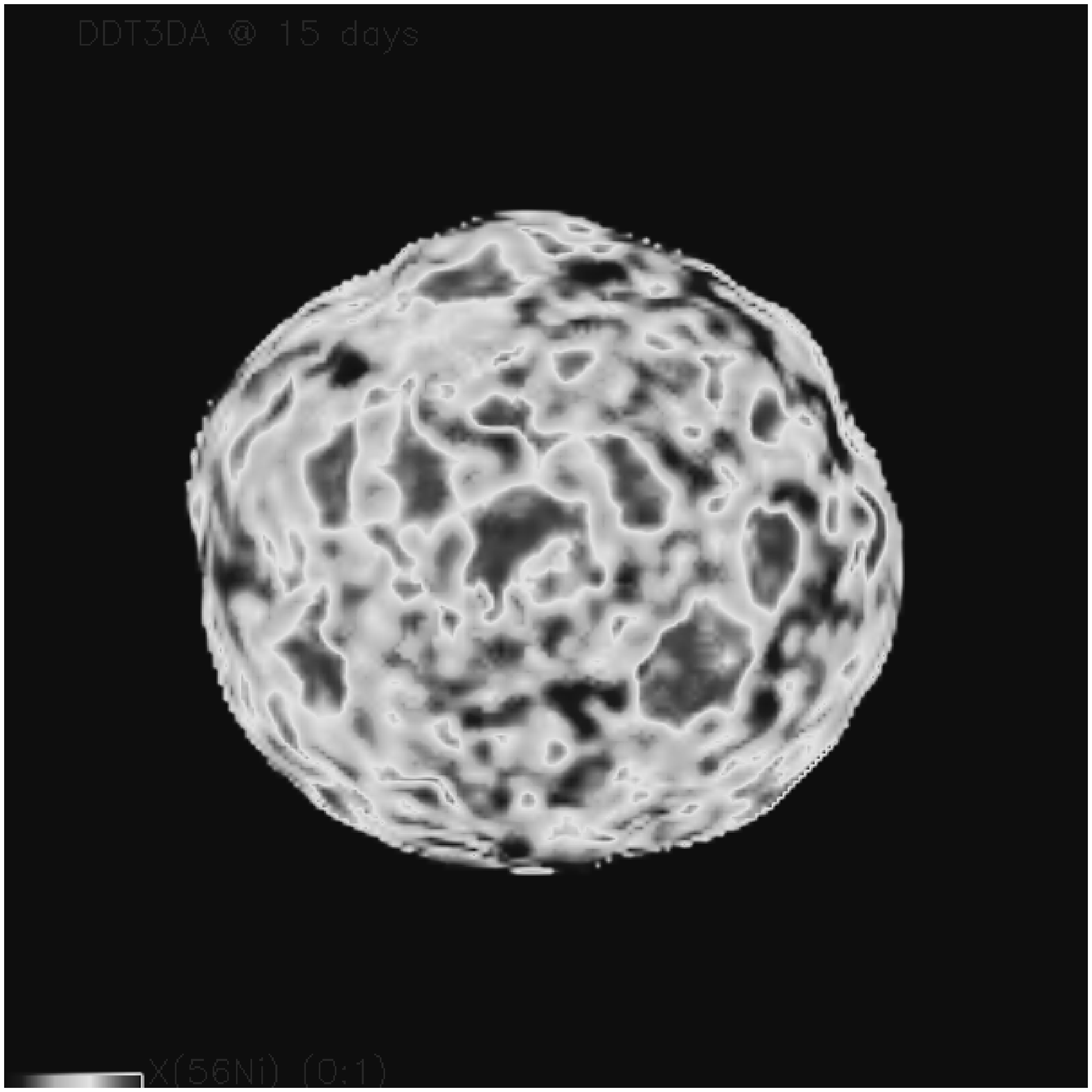}
  \caption{Mass fraction of $^{56}$Ni at the photosphere 15 days after the
  explosion. {\sl Left}: Deflagration model. {\sl Right}: Delayed detonation
  model}
  \label{fig:a}
\end{figure}

In the left panel of Fig.~\ref{fig:a} it is
shown the distribution of $^{56}$Ni at the photosphere for the model starting
from 30 bubbles of the same size \citep{GarciaBravo2004}. The size of the 
clumps is too large to be compatible with the observational limits posed by the
homogeneity of the spectral features of SNIa \citep{Thomas2002}.
It is important to keep in mind that the main properties of 3D deflagrations 
are model independent, as they result from first principles. 
In particular, the deformation of the flame front due to hydrodynamical instabilities
is unavoidable, because the timescale for developing the Rayleigh-Taylor instability
is only a few tenths of a second, i.e. about a factor five lower than the time 
needed by
the deflagration to reach the white dwarf surface. Once the spherical symmetry 
of the flame is lost it is quite difficult to restore it unless a very 
energetic and impulsive phenomenon, like a detonation, is invoked.

In spite of the success achieved by one-dimensional delayed detonation TSN 
models, the physical mechanism responsible for the transition from the 
deflagration to a detonation at the convenient
densities ($\sim10^7$~g\,cm$^{-3}$) is still unknown. Up to now two different 
scenarios have been proposed :
\begin{itemize}
\item A local transition, induced when a fluid element burns with a supersonic phase
velocity when the flame changes from the laminar to the distributed regime 
\citep{Khokhlov1991}.
\item A macroscopic transition, triggered by 
a complex topology of the flame that results in a fuel consumption rate larger 
than that obtained in a supersonic spherical front \citep{WoosleyWeaver1994}.
\end{itemize}
The first mechanism has been studied by \citet{Lisewski2000} who found that the
required mass of the detonator was too large, precluding the formation of
a detonation at the densities of interest. The viability of 
the second mechanism has not been demonstrated so far.
Therefore, delayed detonation calculations in 3D are constrained by the uncertainty
about the transition density, assuming there is any transition at all. In
Table~\ref{tab:a} we show the results of the few 3D models of delayed
detonation that have been computed up to now. The results obtained by different
groups show a larger discrepancy than those derived from pure 3D deflagration 
simulations.
The kinetic energy and the masses of $^{56}$Ni and intermediate mass elements
are, in general, larger than in 3D deflagration models. Nevertheless, the amount of
unburned C-O ejected in the models computed by \citet{GarciaBravo2002} is still 
quite large. In contrast to this situation, the delayed detonation models 
computed by \citet{Gamezo2004} showed that there was no fuel left at the center 
after the passage of the detonation front. The reason
for this apparent discrepancy is the large density of the transition
($>10^8$~g\,cm$^{-3}$) adopted by these
authors. Both groups also obtained different results with respect to the
stratification of the chemical composition in the ejecta. 

The clumps formed during the deflagration phase are destroyed by the
detonation waves. The distribution of $^{56}$Ni 
resulting for the
macroscopic transition delayed detonation model of \citet{GarciaBravo2002} (see
Table~\ref{tab:a}) is shown in the right panel of Fig.~\ref{fig:a}. There, it 
can be
seen how the size of the individual clumps is smaller than in the deflagration
case, making 3D delayed detonations compatible with the spectral homogeneity of
SNIa.

\subsection{New explosion paradigms}

Although most of the {\sl mildly successful} 3D deflagration models 
calculated so far start from a large number of bubbles scattered through the
central region of a white dwarf, nowadays it is not clear how many hot spots can
be present at runaway. Thus, it is interesting to ask what would be the outcome
of the explosion if the initial number of bubbles were small? At first sight, 
one can expect that the energetics of the explosion would be smaller than in the
many bubbles models, probably giving rise to a failed explosion and a pulsation 
of the white dwarf. In this way, the uncertainty about the initial configuration
of the flame has allowed the introduction of two new 
paradigms of TSN, the so-called Pulsating Reverse Detonation (PRD) and the
Gravitationally Confined Detonation (GCD). 

The PRD mechanism of explosion is a byproduct of the simulations of 
deflagrations carried out by
\citet{GarciaBravo2004} starting from 6-7 bubbles. In these simulations, the 
nuclear energy generated during the deflagration phase was
insufficient to unbind the star, as
expected. Due to the ability of hot bubbles to float to
large radii in 3D models, most of the thermal and kinetic energy resided in the outer
parts of the structure, which resulted in an early stabilization of the central
region (mostly made of cold C and O, i.e. fuel) while the outer layers were
still in expansion. A few seconds later, an accretion shock formed at the border
of the central nearly hydrostatic core (whose mass was about $0.9~M_\odot$). 
As a consequence, the temperature
at the border of the core increased to more than $2\times10^9$~K, on a material
composed mainly by fuel but with a non-negligible amount of hot ashes, thus
giving rise to a highly explosive scenario. If a detonation were ignited at this
point, it would probably propagate all the way inwards through the core, burning
most of it and producing an energetic explosion, with a stratified composition. 
A one-dimensional follow-up calculation of the detonation stage produced a
kinetic energy of $0.89\times10^{51}$~erg and $0.35~M_\odot$ of $^{56}$Ni. At
the same time, the amount of unburned C-O was reduced to $0.22~M_\odot$.  

The GCD mechanism of explosion of \citet{Plewa2004} keeps some similarity 
with the PRD, but
this time the runaway starts in a single hot bubble located close to the
center of the white dwarf. Once again, the evolution is dominated by the bubble
motion towards the surface, which hinders a substantial 
propagation of the deflagration front and determines the explosion failure.
However, it is just following this failure of the deflagration, and the
subsequent breakout of the bubble at the surface of the white dwarf, when the
most interesting events take place. The bubble material is then spreaded around
the surface, where it experiences a strong lateral acceleration while remaining 
gravitationally confined to the white dwarf. Finally, the material focuses at 
the
pole opposite to the point of breakout, providing a high compression and
attaining a high temperature ($>2.2\times10^9$~K). The calculations of
\citet{Plewa2004} end at this point, but they claim that a detonation will
probably form at the point of maximum temperature, propagating through
the remaining of the white dwarf and burning it to Fe-group and intermediate mass
elements. It has to be noted that this model was the result of a 2D calculation,
and its results have still to be confirmed by full 3D simulations.

\section{New windows to SNIa: The SN-SNR connection}

The X-ray spectra of supernova remnants originated by SNIa contain important
information regarding the physical mechanism behind the explosions. In the
process of formation of the remnant, the supernova ejecta interact with the
ambient medium surrounding the supernova progenitor, transfer mechanical energy
to it, and are heated through shock waves to a state in which both the ambient
medium and the ejecta emit X-rays. During the initial phase of the remnant
evolution, when the SNR is still young, the emission in the high energy band 
is determined mainly by the properties of the ejecta. In general, the 
X-ray emission can have several components: 
\begin{itemize}
\item A non-thermal continuum, related to the ambient magnetic field,
non-maxwellian populations, etc., 
\item a thermal continuum (bremsstrahlung), sensitive to the local 
state of the plasma, and
\item thermal line emission, sensitive to the chemical abundances, ionization 
state, and thermal state of the plasma.
\end{itemize}
The X-ray line emission from young SNRs provides a convenient way to 
constrain the nucleosynthetic production and energetic properties of the
explosion.

In a recent work, \citet{Badenes2003} showed that the differences in chemical
composition and density profile of SNIa ejecta 
have indeed a deep impact on the thermal X-ray spectra emitted by young
SNRs. Thus, it is possible to use the excellent X-ray spectra of Type Ia
SNRs obtained by X-ray observatories, like {\sl XMM-Newton} and {\sl Chandra}, 
to
constrain Type Ia SN explosion models. Similar approaches were taken in the past
by \citet{DwarkadasChevalier1998} and \citet{Itoh1988}, although these works
either did not include spectral calculations or were limited to the study of a 
particular model (the W7 model). \citet{BlinnikovSorokina2002} have
undertaken a similar enterprise, whose results are just beginning to appear in
the literature. More recently, \citet{Badenes2004} has compared the spectra 
predicted by more than
400 supernova remnant models (generated combining 26 SNIa explosion models in 1D
and 3D, incluing all the explosion mechanisms currently under debate, with different
assumptions about the physical state of the ambient medium) with the spectra
of well-known remnants, like the Tycho SNR. 

\begin{figure}
  \includegraphics[height=.33\textheight]{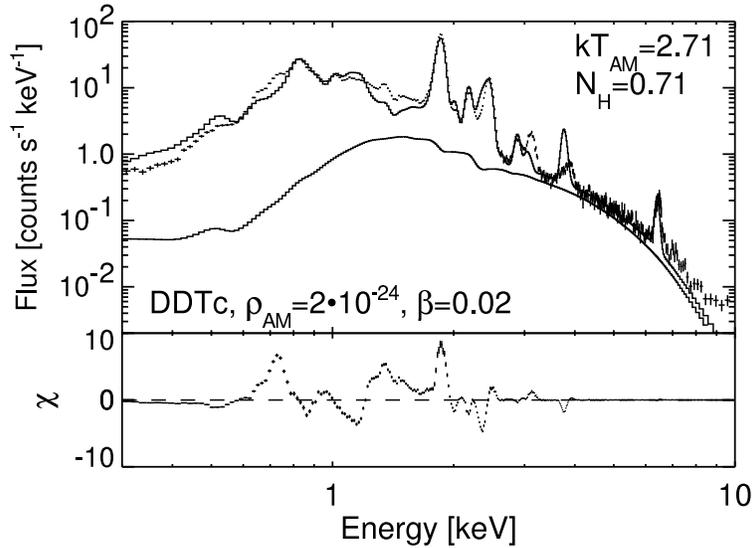}
  \caption{The best fit to the {\sl XMM-Newton} spectrum of the Tycho SNR (solid
  line) is compared to the data (points). The spectrum emitted by the shocked
  interstellar medium is also shown (featureless solid curve). See
  \citet{Badenes2004} for details}
  \label{fig:b}
\end{figure}

The Tycho SNR, which is the best candidate for a Type Ia remnant, has been
extensively studied with both {\sl Chandra} and {\sl XMM-Newton}, providing
high-quality spectra with precise determinations of the flux and energy centroid
of the spectral features produced by each element. One of the main properties of
the X-ray spectra of the Tycho remnant is that the emission lines due to Fe and
Si (and other intermediate mass elements, like S and Ca) are produced under
quite different thermal conditions. Through a careful
analysis, \citet{Badenes2004} has been able to prove that most of 
the explosion paradigms are incompatible with the spectra of Tycho. The best 
approximation to the
X-ray spectrum of the Tycho SNR is obtained with a mildly-energetic ($K =
1.2\times10^{51}$~erg) 1D (i.e. chemically stratified) delayed detonation model,
which synthesizes $0.74~M_\odot$ of $^{56}$Ni (see Fig.~\ref{fig:b}). 
It was shown in this work that, due to the absence of chemical stratification, 
the SNR models produced by 3D 
deflagrations and 3D delayed detonations were characterized by the homogeneity of
the ionization and thermal properties of all the chemical elements. As we have 
explained before, such
homogeneity is incompatible with the physical properties derived from the
X-ray spectra of the Tycho SNR and other candidate Type Ia SNRs. It will be 
interesting to see whether the spectra 
predicted by the new 3D
explosion paradigms of SNIa will retain these characteristics or, on the
contrary, will become more similar to the ones corresponding to 1D explosion 
models.

\section{Summary}

Since \citet{HoyleFowler1960} proposed the white dwarf scenario for Type Ia SNe, the ideas
about the way the star explodes have evolved. The 70's were the epoch of pure
detonations. The 80's witnessed a flourishing of the deflagrations, mainly thanks
to the popular W7 model. The 90's knew about delayed detonations in its various flavors.
Nowadays, at the beginning of the 21st century, the future of TSN modelling resides most 
probably on new paradigms, like PRD and GCD. 

Although, in the near future, the analysis of Type Ia supernovae will continue being based
predominantly on optical observations, the realm of high energies is going
to play an 
increasingly important role in the understanding of these objects. We have
discussed some recent progress on the applications of X-ray 
spectra from young supernova remnants to the determination of the explosion
mechanism. Future observations of gamma-rays from SNIa will
allow a much more in depth knowledge of the amount of radioactive nuclei 
produced in the explosion, its distribution throughout the ejecta, and the
eventual influence of the interaction with a secondary star in a binary system
\citep[see, e.g.][]{Isern2004}.


\begin{theacknowledgments}
This research has been partialy supported by the CIRIT, the MEC programs
AYA2002-04094-C03-01/02 and AYA2004-06290-c02-01/02, and by the EU FEDER 
funds.             
\end{theacknowledgments}


\bibliographystyle{aipproc}   

\bibliography{bravo}

\IfFileExists{\jobname.bbl}{}
 {\typeout{}
  \typeout{******************************************}
  \typeout{** Please run "bibtex \jobname" to optain}
  \typeout{** the bibliography and then re-run LaTeX}
  \typeout{** twice to fix the references!}
  \typeout{******************************************}
  \typeout{}
 }

\end{document}